\RequirePackage{lineno}
\documentclass[a4paper,11pt,epsfig]{article}
\pdfoutput=1 

\usepackage{jheppub} 

\usepackage[T1]{fontenc} 
\usepackage{graphicx,color,dcolumn,booktabs,bm}
\usepackage{longtable,lscape}
\usepackage{amsmath}
\usepackage{indentfirst}
\usepackage{epsfig}
\usepackage{epstopdf}   
\usepackage{slashed}  
\usepackage{color}
\usepackage{multirow}
\usepackage{soul}  
\usepackage{graphicx,color,dcolumn,booktabs,bm}
\usepackage{verbatim}
\usepackage{orcidlink}

\newcommand{\BR}{{\cal B}}

\newcommand{\beq}{\begin{equation}}
\newcommand{\eeq}{\end{equation}}
\newcommand{\bitm}{\begin{itemize}}
\newcommand{\eitm}{\end{itemize}}

\newcommand{\omec}{\Omega^0_c}
\newcommand{\octxipi}{\Omega^0_c\to\Xi^-\pi^+}
\newcommand{\octxik}{\Omega^0_c\to\Xi^-K^+}
\newcommand{\octopi}{\Omega^0_c\to\Omega^-\pi^+}
\newcommand{\octok}{\Omega^0_c\to\Omega^-K^+}

\renewcommand\tabcolsep{2.0pt}

\title{\boldmath Evidence for the singly Cabibbo-suppressed decay $\octxipi$ and search for $\octxik$ and $\Omega^-K^+$ decays at Belle}

\vspace{20cm}
\vspace{20cm}

\collaboration{The Belle Collaboration}
  \author{X.~Han\,\orcidlink{0000-0003-1656-9413},} 
  \author{S.~Jia\,\orcidlink{0000-0001-8176-8545},} 
  \author{L.~Yuan\,\orcidlink{0000-0002-6719-5397},} 
  \author{C.~P.~Shen\,\orcidlink{0000-0002-9012-4618},} 
  \author{I.~Adachi\,\orcidlink{0000-0003-2287-0173},} 
  \author{J.~K.~Ahn\,\orcidlink{0000-0002-5795-2243},} 
  \author{H.~Aihara\,\orcidlink{0000-0002-1907-5964},} 
  \author{D.~M.~Asner\,\orcidlink{0000-0002-1586-5790},} 
  \author{T.~Aushev\,\orcidlink{0000-0002-6347-7055},} 
  \author{R.~Ayad\,\orcidlink{0000-0003-3466-9290},} 
% ...

\abstract{Using a data sample of 980 fb$^{-1}$ collected with the Belle detector at the KEKB asymmetric-energy $e^+e^-$ collider, we study for the first time the singly Cabibbo-suppressed decays $\Omega^0_c\to\Xi^{-}\pi^{+}$ and $\Omega^-K^+$ and the doubly Cabibbo-suppressed decay $\Omega^0_c\to \Xi^- K^{+}$.
Evidence for an $\Omega^0_c$ signal in the $\Omega^0_c\to\Xi^-\pi^+$ mode is reported with a significance of $4.5\sigma$ including systematic uncertainties.
The ratio of branching fractions to the normalization mode $\Omega_c^0\to \Omega^-\pi^+$ is measured to be 
$$\BR(\Omega_{c}^{0} \to \Xi^{-} \pi ^{+} )/\BR(\Omega_c^0\to \Omega^-\pi^+)=0.253\pm 0.052({\rm stat.})\pm 0.030({\rm syst.}).$$
No significant signals of $\octxik$ and $\Omega^-K^+$ modes are found.
The upper limits at $90\%$ confidence level on ratios of branching fractions are determined to be
$$\BR(\Omega_{c}^{0} \to \Xi^{-} K ^{+})/\BR(\Omega_c^0\to \Omega^-\pi^+) < 0.070$$
and
$$\BR(\Omega_{c}^{0} \to \Omega^{-} K ^{+})/\BR(\Omega_c^0\to \Omega^-\pi^+)  < 0.29.$$
}


\begin{document} 
\maketitle
\flushbottom

\section{Introduction}
In comparison with the other weakly-decaying singly-charmed baryons states, the $\Lambda^+_c$, $ \Xi^0_c $, and $ \Xi^+_c $, our knowledge of the $\Omega_c^0$ state is limited~\cite{PDG}. 
There were a few experiments that studied the $\Omega_c^0$ baryon decays over the past two decades.
There are no measurements of the absolute branching fractions of the $ \omec $ decays,
but some measurements of the branching fraction ratios for $ \omec $ decay modes with respect to the 
normalization mode $ \Omega ^- \pi ^+ $ have been made~\cite{PDG}.
CLEO reported $\omec$ decays using five decay modes ($ \Omega^- \pi^+,~\Omega^- \pi^+ \pi^0,~\Xi^- K^- \pi^+ \pi^+,~\Xi^0 K^- \pi^+ $, and $\Omega^- \pi^+ \pi^+ \pi^-$)~\cite{cleo5}, and observed the semileptonic decay of $ \omec\to \Omega ^- e^+ \nu_e $~\cite{cleo1}.
BABAR reported $\omec$ decays and measured ratios of branching fractions
for four final states ($\Omega^-\pi^+$, $\Omega^- \pi^+ \pi^0$, $\Omega^-\pi^+\pi^+\pi^-$, and $\Xi^-K^-\pi^+\pi^+$)~\cite{babar4}.
Belle measured the $\omec$ decay of $\omec\to \Omega ^- \pi ^+ $~\cite{belle1}, and reported ratios of branching fractions for $\Omega^-\pi^+\pi^0$, $\Omega^- \pi^+ \pi^- \pi^+$, $\Xi^- K^- \pi^+ \pi^+$, $\Xi^0 K^- \pi^+$,
$\Xi^-\bar{K}^0 \pi^+$, $\Xi^0\bar{K}^0$, and $ \Lambda \bar{K}^0\bar{K}^0 $~\cite{bellen}. Very recently, the branching fraction ratios for semileptonic decays of $\omec\to \Omega^-\ell^+\nu_{\ell}$ ($\ell$ = $e$ or $\mu$) have been measured by Belle with improved precision~\cite{2112.10367}. 

The $\Omega_{c}^{0}$, which has $J^{P}=\left(\frac{1}{2}\right)^{+}$, is the heaviest singly-charmed hadron that decays weakly.
The quark content of the $\Omega_{c}^{0}$  is  $c\{s s\}$, where the $s s$ 
pair is in a symmetric state.~The theoretical studies of hadronic weak decays of the $ \Omega _{c}  $ baryon have a long history over several decades~\cite{2109.01216}.
Various methods have been developed to describe the nonfactorizable contributions 
which play an important role in the hadronic decays.
There are many theoretical models that have predicted
the mass and branching fractions of the  $\Omega_{c}^{0}$~\cite{659,3836,4188,5632,114008,025205,074011,093101,094033}.~However, the range of these predictions is rather wide. For the singly Cabibbo-suppressed (SCS) and doubly Cabibbo-suppressed (DCS) decays of $\octxipi$ and $\octxik$, the branching fractions have been calculated using the light-front quark model (LFQM)~\cite{093101}, pole model~\cite{094033}, and current algebra (CA)~\cite{094033}, and these are listed in table~\ref{tabrefs}. The studies of $\octxipi$ and $\octxik$ are crucial to test the theoretical models by comparing the measured branching fractions and corresponding theoretical predictions~\cite{093101,094033}.
No prediction is available for $\omec\to\Omega^- K^+$.

\begin{table}[htpb]
\centering
\caption{Predicted ratios of branching fractions for $\octxipi$ and $\octxik$ using LFQM~\cite{093101}, pole model~\cite{094033}, and CA~\cite{094033}. The branching fraction of reference mode $ \octopi $ is $ 9\% $ according to Ref.~\cite{094033}.}
\vspace{0.2cm}
\label{tabrefs}
\begin{tabular}{c c c}
\hline\hline
~~~~~Decay modes~~~~~ & ~~~~~LFQM~\cite{093101}~~~~~ & ~~~~~pole model and CA~\cite{094033}~~~~~ \\\hline
$\octxipi$ & 1.96 $\times$ 10$^{-3}$ & 1.04 $\times$ 10$^{-1}$ \\
$\octxik$ & 1.74 $\times$ 10$^{-4}$ & 1.06 $\times$ 10$^{-2}$ \\
\hline\hline
\end{tabular}                                              
\end{table}

In this article, we use a data sample of 980 fb$^{-1}$ collected by the Belle detector to study for the first time the SCS modes $\octxipi$ and $\octok$, and the DCS mode $\octxik$.
The $\Xi^-$ and $\Omega^-$ are reconstructed from $\Lambda\pi^-$ and $\Lambda K^-$, followed by $\Lambda\to p \pi^-$.
Throughout this analysis, for any given mode, the corresponding charge-conjugate mode is also implied. We report the ratios of branching fractions to the normalization mode $\Omega_c^0\to \Omega^-\pi^+$. 

\section{Data sample and the Belle Detector}

This measurement is based on data recorded at or near the $\Upsilon(1S)$, $\Upsilon(2S)$, $\Upsilon(3S)$, $\Upsilon(4S)$, and $\Upsilon(5S)$ resonances by the Belle detector~\cite{Belle1,Belle2} at the KEKB asymmetric-energy $e^+e^-$ collider~\cite{KEKB1,KEKB2}. The data sample corresponds to a total integrated luminosity of 980 fb$^{-1}$~\cite{Belle2}. The Belle detector is a large-solid-angle magnetic spectrometer that consists of a silicon vertex detector, a 50-layer central drift chamber (CDC), an array of aerogel threshold Cherenkov counters (ACC), a barrel-like arrangement of time-of-flight scintillation counters (TOF), and an electromagnetic calorimeter comprised of CsI(Tl) crystals (ECL) located inside a superconducting solenoid coil that provides a 1.5 T magnetic field. An iron flux-return yoke instrumented with resistive plate chambers located outside the coil is used to detect $K^0_L$ mesons and identify muons. A detailed description of the Belle detector can be found in Refs.~\cite{Belle1,Belle2}. 

Monte Carlo (MC) signal samples are generated using {\sc evtgen}~\cite{EVTGEN} to optimize the signal selection criteria and calculate the signal reconstruction efficiency; $e^+e^-\to c\bar c$ events are simulated using {\sc pythia}~\cite{PYTHIA} model, and $\omec\to \Xi^-  \pi^{+}/\Xi^- K^{+}/ \Omega ^{-} K^+$ decays are generated with a phase space model. 
The EvtGen generator widely used in Belle experiment can produce a simulated event stream realistic enough to be essentially indistinguishable from real data~\cite{3026}.
Events containing the reference mode, $\Omega_c^0\to\Omega^-\pi^+$, are produced with its known angular distribution~\cite{Babar}.
The effect of final-state radiation is taken into account in the simulation using the {\sc photos}~\cite{291} package. The simulated events are processed with a detector simulation based on {\sc geant3}~\cite{geant3}. 
The simulated events 
Generic MC samples, i.e. $B$ ($=B^+$, $B^0$, or $B^{(*)}_s$) decays and $e^+e^-\to q\bar q$ ($q$ = $u$, $d$, $s$, $c$) at $\sqrt{s}$ = 10.52, 10.58, and 10.867 GeV, having four times integrated luminosity as real data, are used to evaluate possible peaking backgrounds, optimize selection criteria, and study the signal and background invariant mass shapes.

\section{Selection criteria}

Except for the charged tracks from the relatively long-lived $\Lambda$, $\Xi^-$, and $\Omega^-$ decays,
all charged tracks are required to originate from the vicinity of the interaction point (IP).
The impact parameters perpendicular to (${\rm d}r$) and along the beam direction ($|{\rm d}z|$) with respect to the IP are required to be less than 0.2 cm and 1 cm, respectively.
For the particle identification (PID) of a track, information from different detector subsystems, including specific ionization in the CDC, time measurement in the TOF, and the response of the ACC, is combined to form a likelihood ${\cal L}_i$~\cite{PID} for particle species $i$. Tracks with $R_K={\cal L}_K/({\cal L}_K+{\cal L}_\pi)$ $<$ 0.4 are identified as pions with an efficiency of $95.5\%$, while $8.0\%$ of kaons are misidentified as pions; tracks with $R_K$ $>$ 0.6 are identified as kaons with an efficiency of $89.4\%$, while $4.2\%$ of pions are misidentified as kaons.
The PID (misidentified) efficiency is evaluated using the signal MC data sample, where the PID of the tracks are known. It is defined as the number of tracks selected under a PID hypothesis divided by the number of tracks before any PID requirements with the same (opposite) PID hypothesis.

The $\Lambda$ candidates are reconstructed via $p\pi^-$ pairs.
The distance of the $\Lambda$ decay vertex with respect to the IP is greater than 0.35 cm with an efficiency of 99.5\%.
For $\Xi^-\to\Lambda\pi^-$, the vertex formed from the $\Lambda$ and $\pi^-$ is required to be at least 0.35 cm from the IP with an efficiency of $ 94.1\% $, and to be a shorter distance from the IP than the $\Lambda$ decay vertex~\cite{052003,032006}. 
These efficiencies are obtained by comparing the numbers of MC signal events with and without applying these selection requirements.
For $\Omega^-\to \Lambda K^-$, the flight directions of $\Lambda$ and $\Omega^-$ candidates, which are reconstructed from their production and decay vertices after performing vertex and mass constraint fits to the full decay chain, are required to be within five degrees of their momentum directions in both 3D space and the plane perpendicular to the z-axis in the laboratory frame~\cite{2112.10367}, where the z-axis is opposite to the $e^+$ beam direction.
The values of $ \chi^2 $ from vertex fits are required to be less than 20 optimized by maximizing the
figure-of-merit FoM = $N_{\text{sig}}/\sqrt{N_{\text{sig}}+N_{\text{bkg}}}$, where $N_{\text{sig}}$ is the number of expected $\Omega_c^0$ signal events using MC simulations assuming $\BR(\Omega^0_c\to\Xi^-\pi^+)/\BR(\Omega^0_c\to\Omega^-\pi^+)$ and $\BR(\Omega^0_c\to\Xi^-K^+)/\BR(\Omega^0_c\to\Omega^-\pi^+)$ are $1.04\times 10^{-1}$ and $ 1.06\times 10^{-2} $, respectively, from table~\ref{tabrefs}, and $N_{\text{bkg}}$ is the number of estimated background events in the $\Omega_c^0$ signal region using normalized generic MC samples.
The $\Xi^-\pi^+$, $ \Xi^-K^+$, $\Omega^-K^+$, and $\Omega^-\pi^+$ candidates are combined to 
form an $\omec$ candidate and its daughter tracks fitted to a common vertex. 
For the reference mode of $\Omega^0_c\to\Omega^-\pi^+$, selection criteria for signal candidates are the same as those applied in Ref.~\cite{2112.10367} except for the $x_p$ requirement (see below).

To reduce combinatorial backgrounds, especially from $B$ meson decays, the scaled momentum $x_p = p^*/p^*_{\rm max}$ is required to 
be greater than 0.65. Here, $p^*$ is the momentum of $\omec$ in the $e^+e^-$ center-of-mass (C.M.) frame, and $p^*_{\rm max}$ = $\sqrt{E^2_{\rm beam}-M_{\omec}^2c^4}/c$ is the maximum momentum, where $E_{\rm beam}$ is the beam energy in the C.M. frame and $M_{\omec}$ is the invariant mass of $\omec$.
The $x_p$ requirement is optimized in $\Omega^0_c\to\Xi^-\pi^+$ by maximizing the FoM.

\section{Branching fraction ratios of $\octxipi$, $\Xi^-K^+$, and $\Omega^-K^+$}

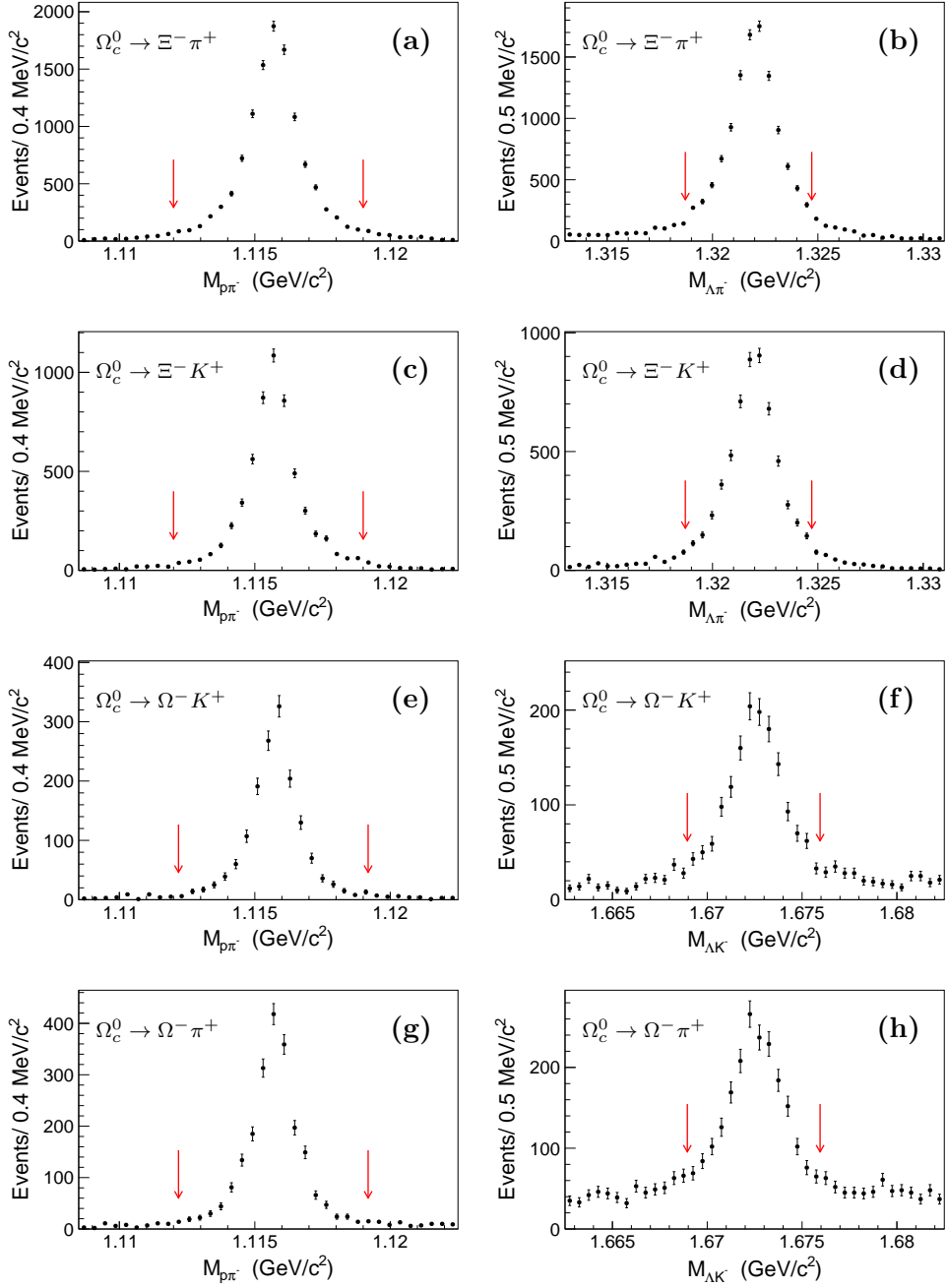
\begin{figure}[htpb]
  \centering
  \includegraphics[width=65mm]{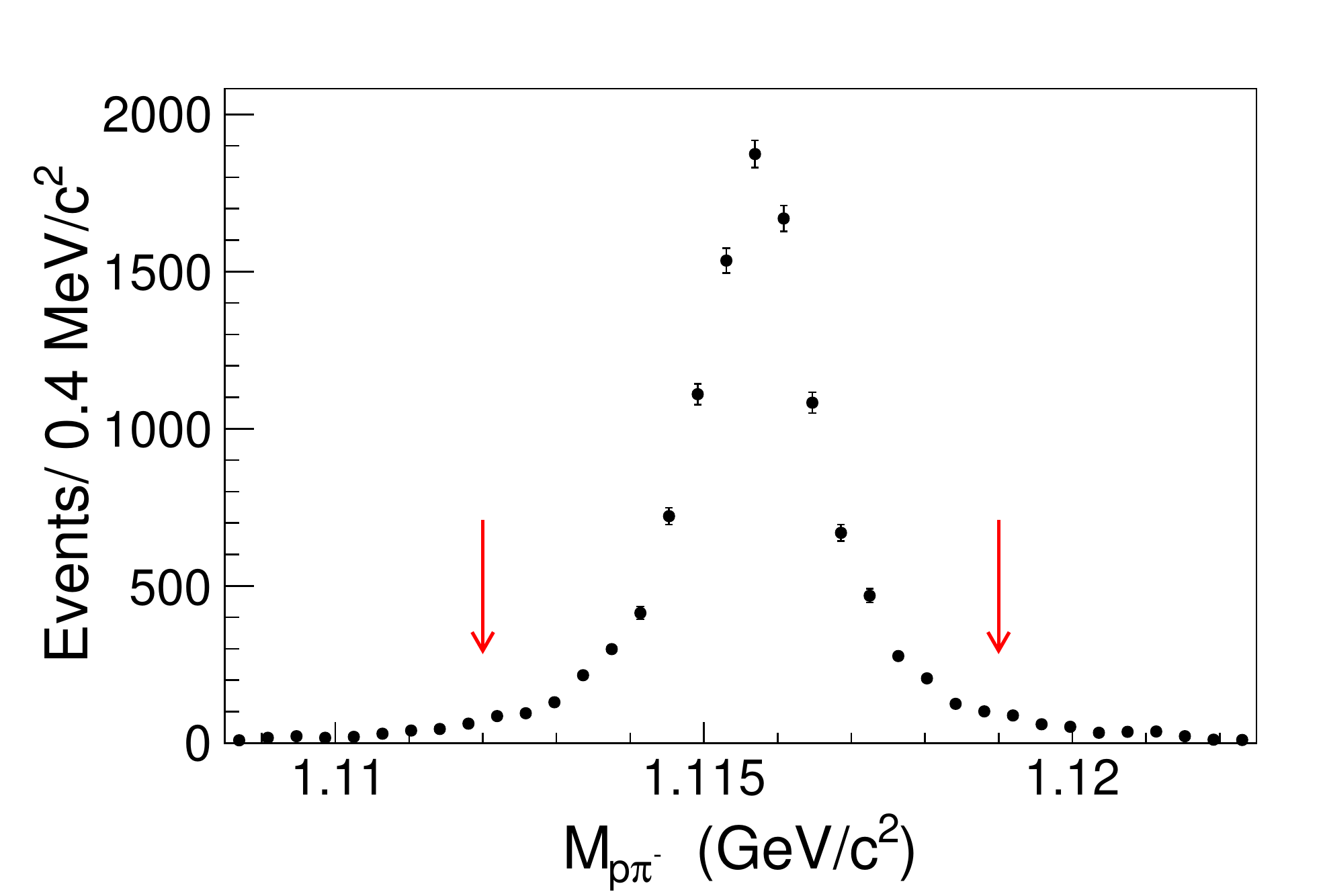}
  \put( -35, 95){\bf (a)}
  \put(-147, 95){\scriptsize $ \octxipi $ }
  \includegraphics[width=65mm]{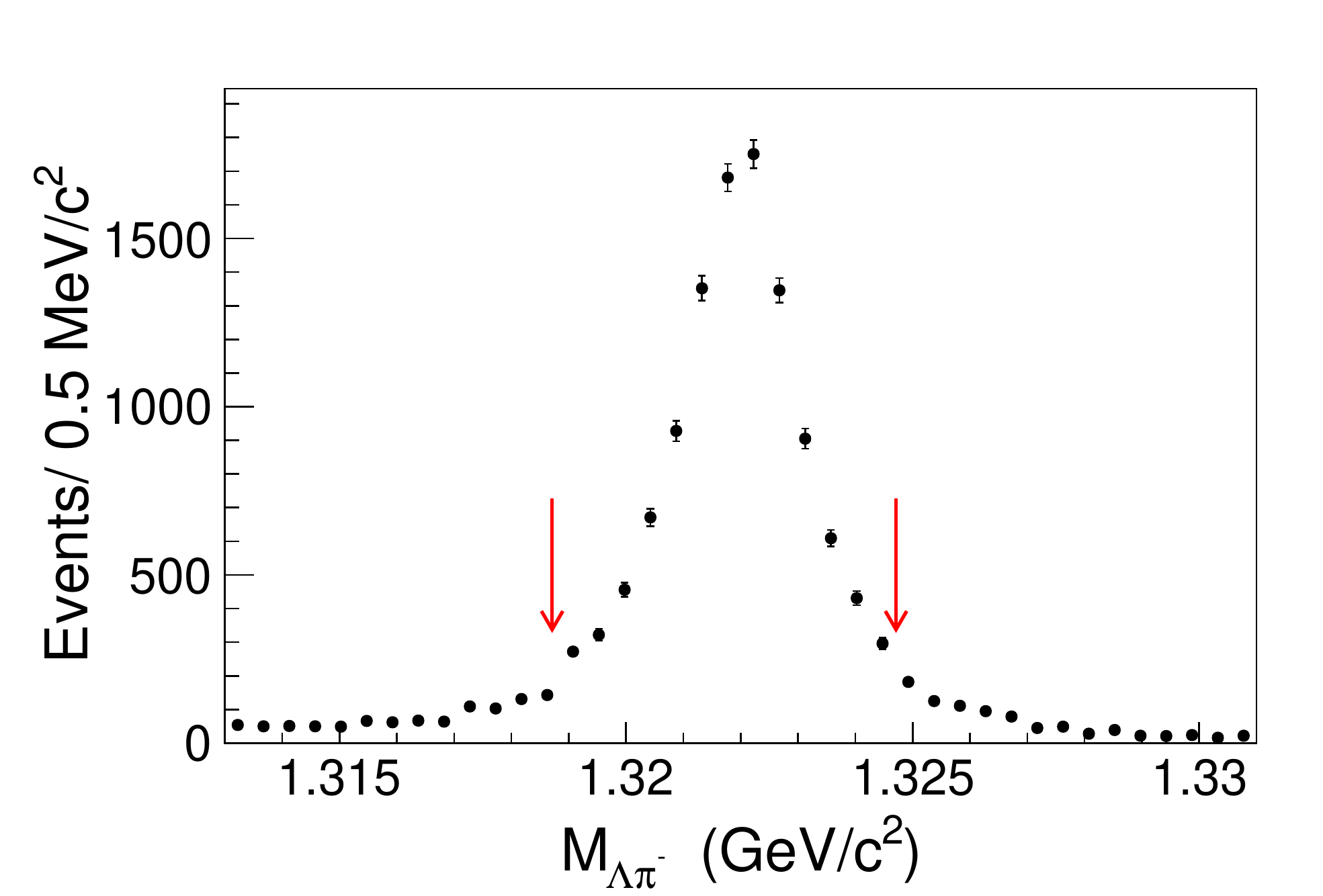}
  \put(-35, 95){\bf (b)}
  \put(-147, 95){\scriptsize $ \octxipi $ }

  \includegraphics[width=65mm]{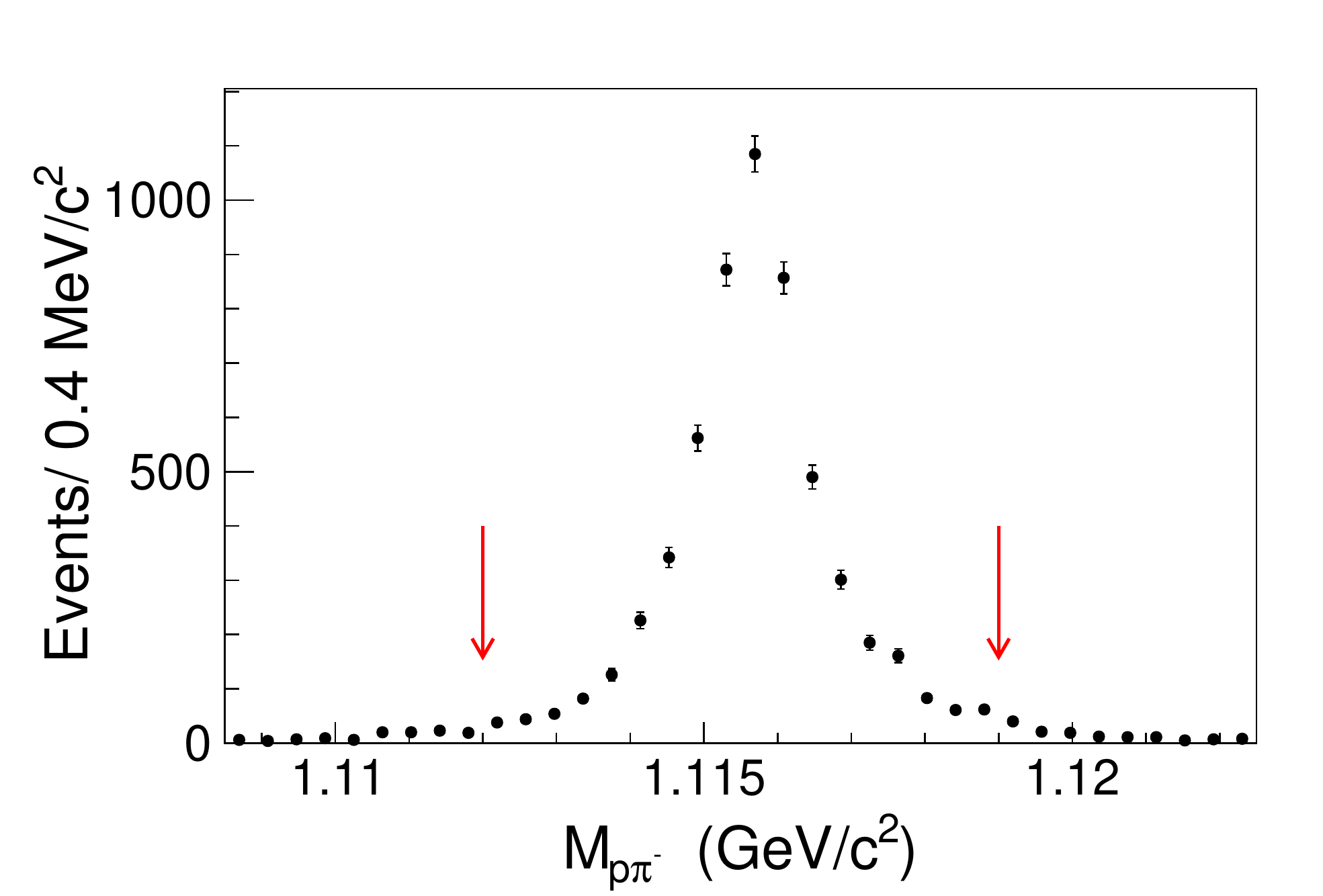}
  \put(-35, 95){\bf (c)}
  \put(-147, 95){\scriptsize $ \octxik $ }
  \includegraphics[width=65mm]{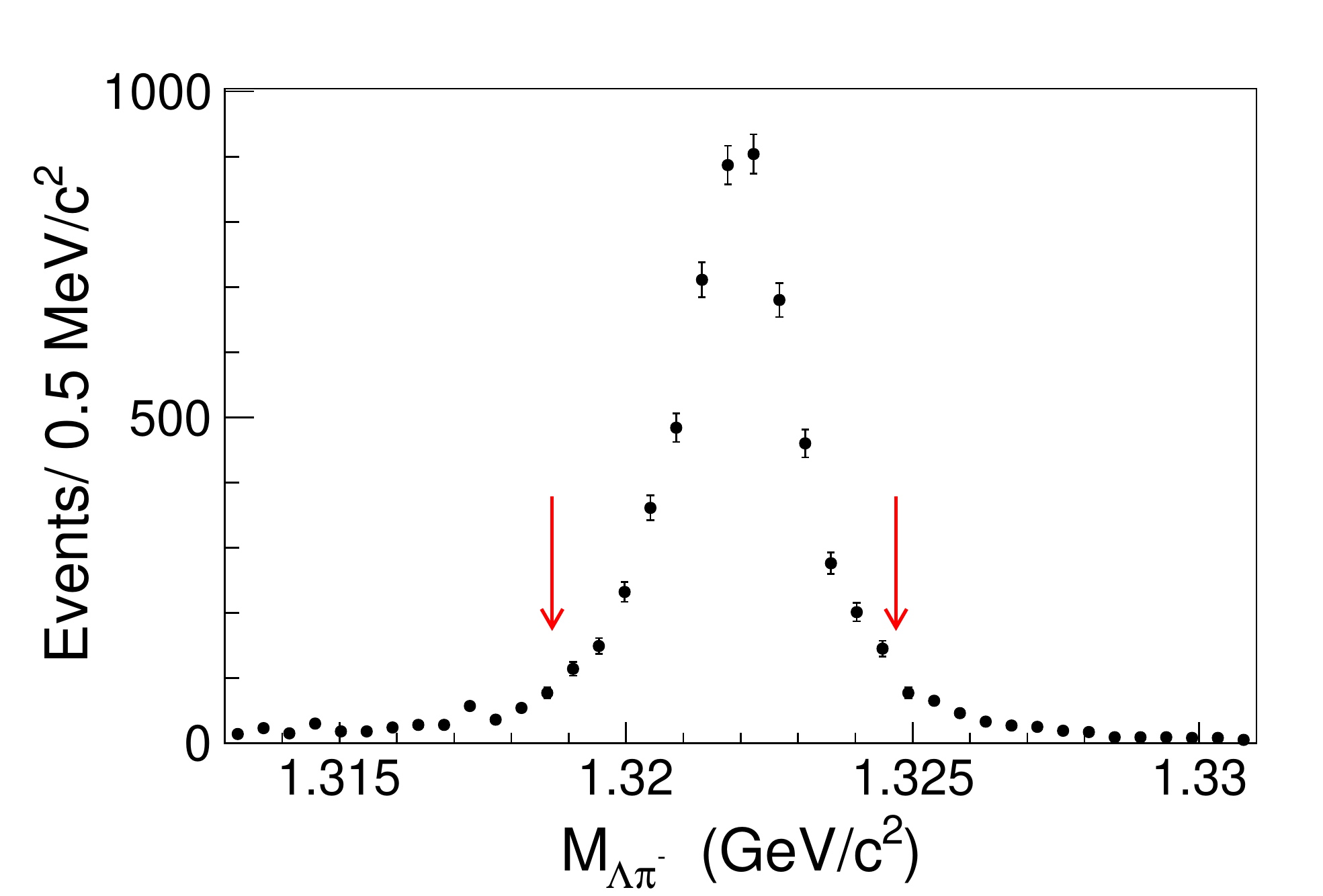}
  \put(-35, 95){\bf (d)}
  \put(-147, 95){\scriptsize $ \octxik $ }

  \includegraphics[width=65mm]{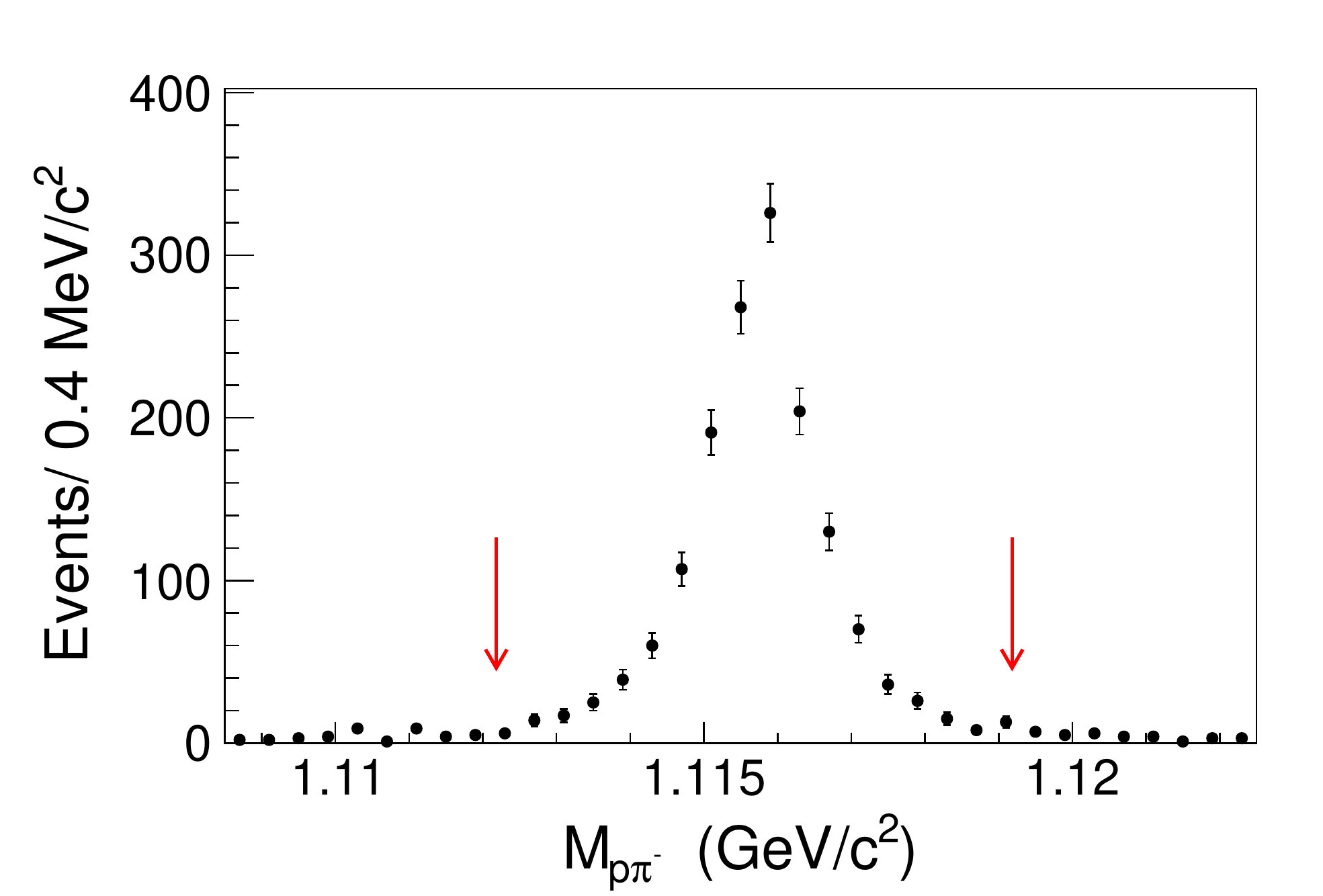}
  \put(-35, 95){\bf (e)}
  \put(-147, 95){\scriptsize $ \octok $ }
  \includegraphics[width=65mm]{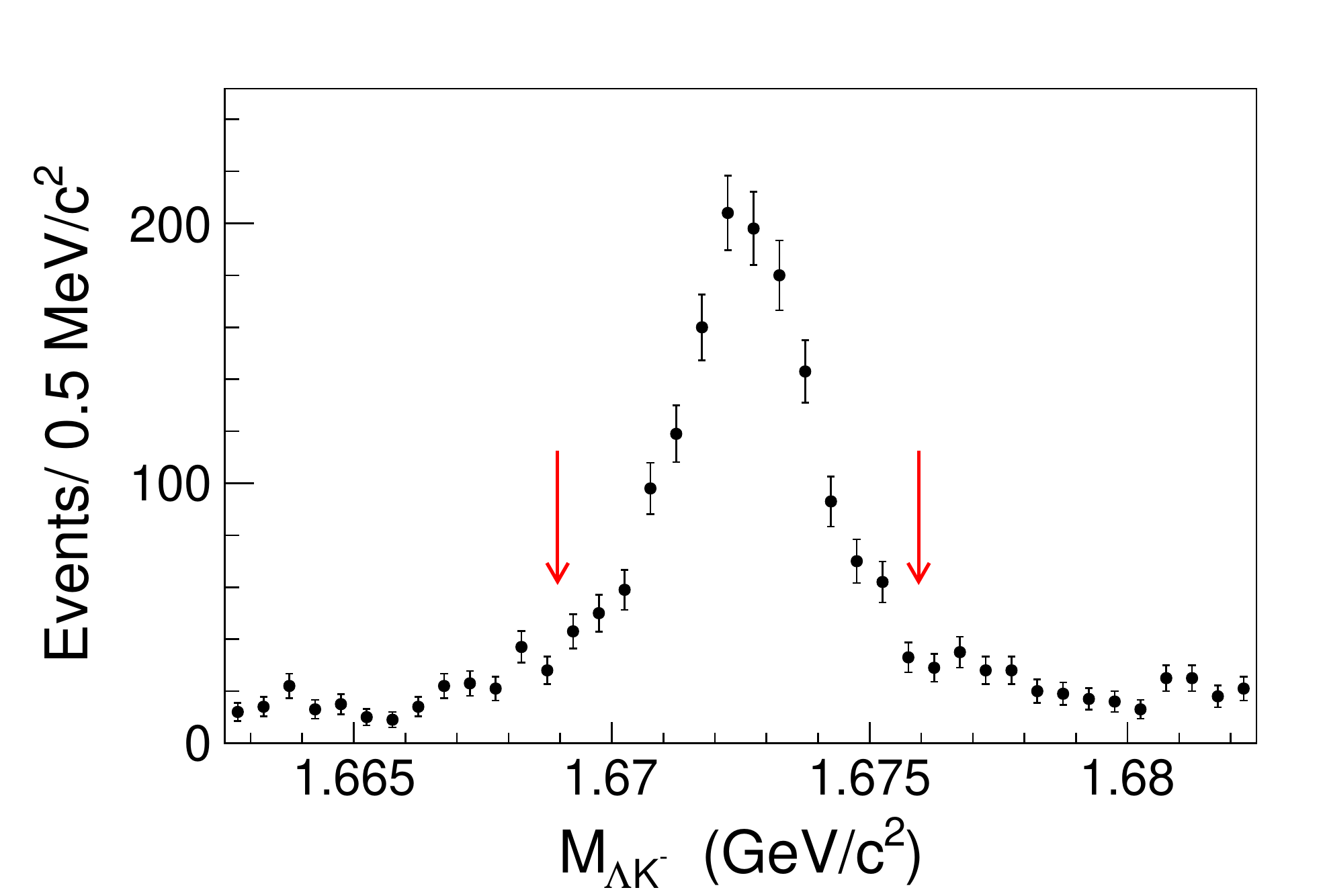}
  \put(-35, 95){\bf (f)}
  \put(-147, 95){\scriptsize $ \octok $ }

  \includegraphics[width=65mm]{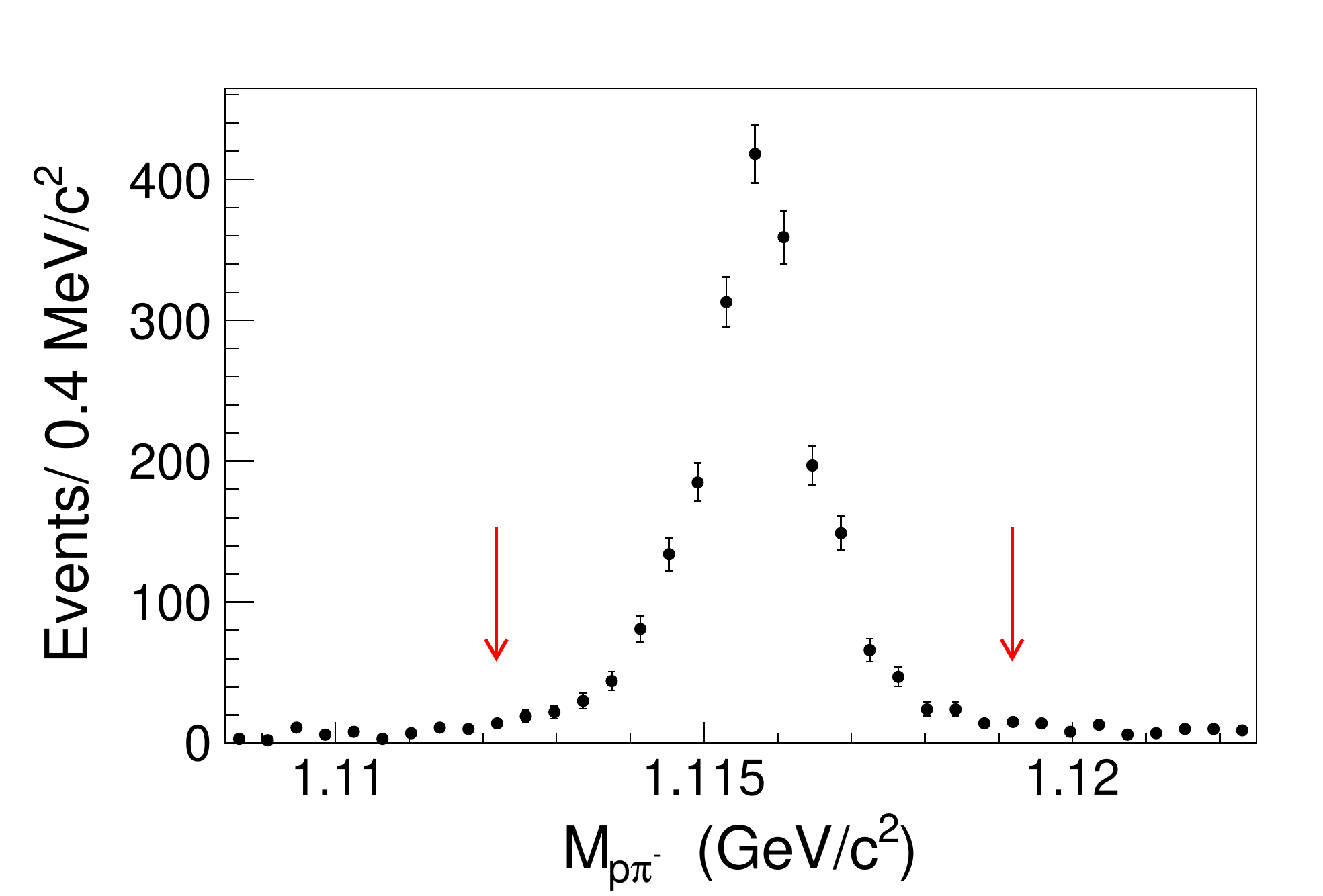}
  \put(-35, 95){\bf (g)}
  \put(-147, 95){\scriptsize $ \octopi $ }
  \includegraphics[width=65mm]{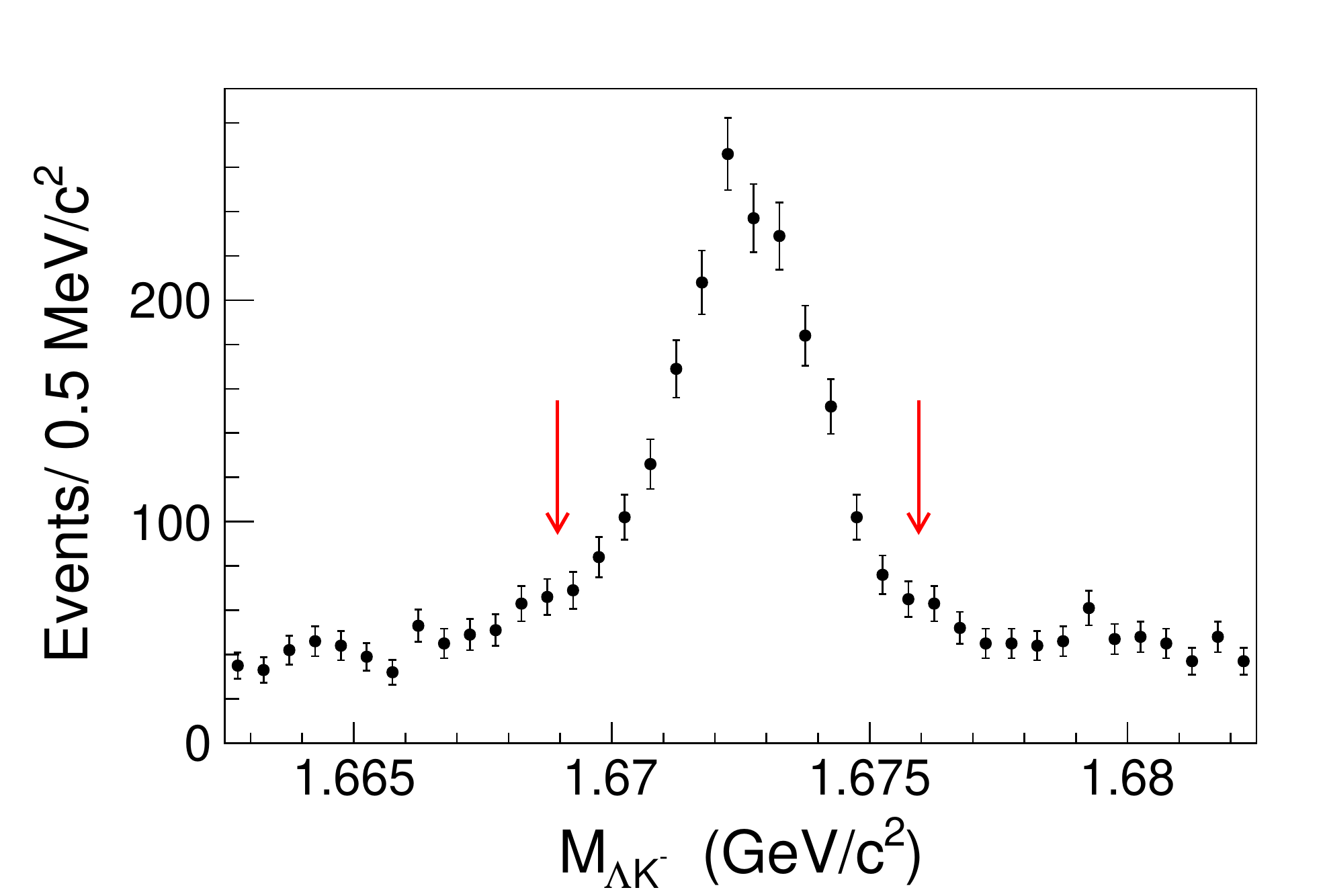}
  \put(-35, 95){\bf (h)}
  \put(-147, 95){\scriptsize $ \octopi $ }
  \caption{The left column shows the $\Lambda$ candidates in (a) $\Omega_c^0 \to \Xi^- \pi^+$, (c) $\Xi^- K^+$, (e) $\Omega^- K^+$, and (g) $\Omega^- \pi^+$ decays. 
In the right column the invariant mass of $\Xi^-$ candidates (in (b) $\Omega_c^0 \to \Xi^- \pi^+$ and (d) $\Xi^- K^+$ decays) and $\Omega^-$ candidates (in (f) $\Omega_c^0 \to \Omega^- K^+$ and (h) $\Omega^- \pi^+$ decays) are shown.
  The red arrows show the required $\Lambda$, $\Xi^-$ and $ \Omega ^- $  signal regions.
  }
  \label{fig:data1}
\end{figure}

After applying the above requirements, the invariant mass distributions of $p\pi^-$, $\Lambda\pi^-$, and $\Lambda K^-$ from data samples for $ \octxipi $, $ \Xi^-K^+ $, $ \Omega^- K^+ $ and reference mode $ \octopi $ are shown in figure~\ref{fig:data1}. The $\Lambda$, $\Xi^-$, and $\Omega^-$ signals are clearly visible in all the invariant mass distributions. We define $\Lambda$, $\Xi^-$, and $\Omega^-$ signal regions as $|M(p\pi^-) - m_{\Lambda}|$ $<$ 3.5~MeV/$c^2$, $|M(\Lambda\pi^-) - m_{\Xi^-}|$ $<$ 3 MeV/$c^2$, and $|M(\Lambda K^-) - m_{\Omega^-}|$ $<$ 3.5 MeV/$c^2$, corresponding to the efficiencies of $97\%$, $92\%$, and $97\%$, respectively, based on MC simulations. The $ \Xi^- $ signal region is optimized in $ \octxipi $ and $ \Xi^-K^+ $ modes by maximizing the FoM. The efficiencies of the signal regions are the same for signal channels and the reference mode.
Hereinafter, $M$ represents a reconstructed invariant mass and $m_i$ denotes the nominal mass of particle $i$~\cite{PDG}.

\begin{figure}[htpb]
  \centering
  \includegraphics[width=75mm]{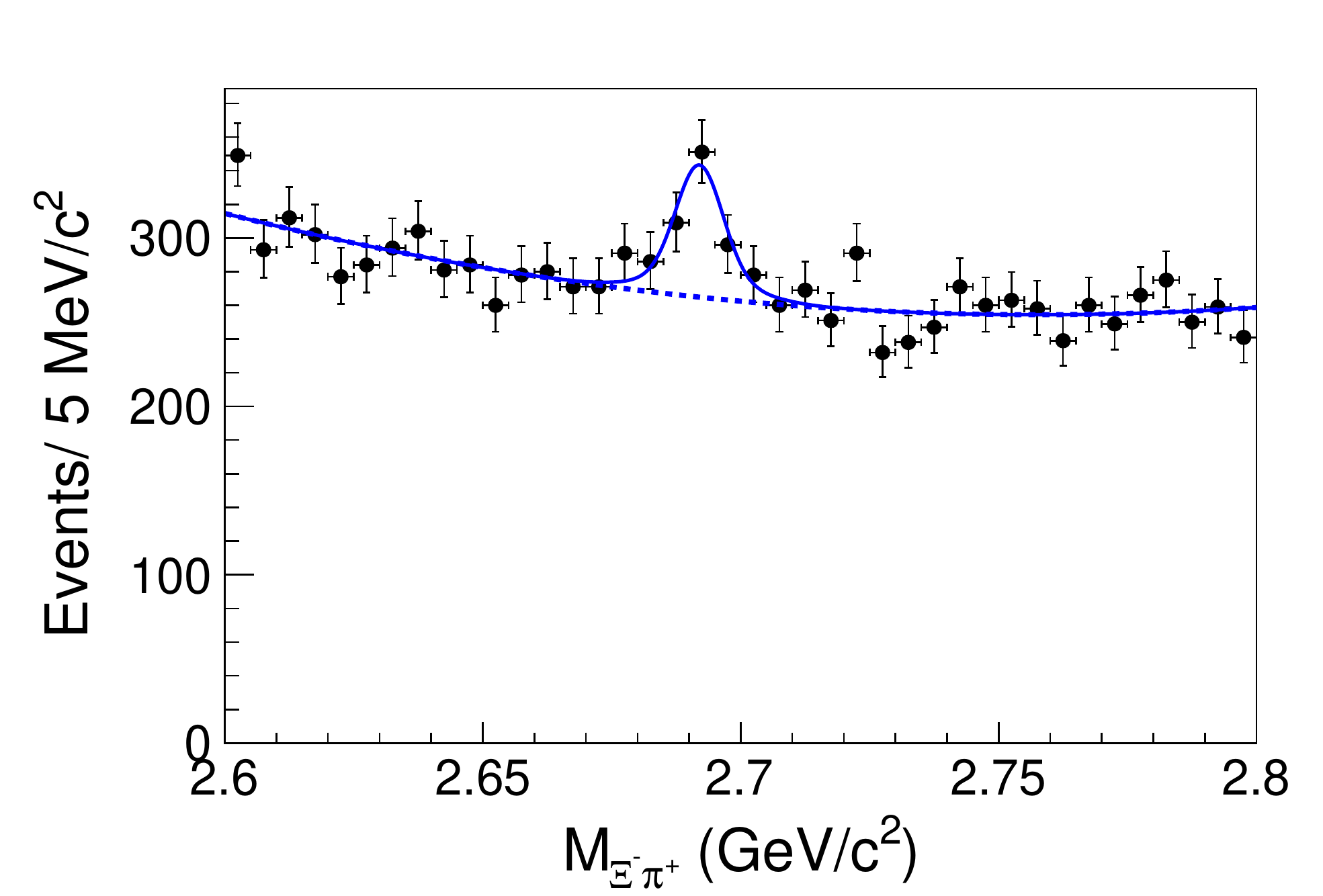}
  \put(-35, 115){\bf (a)}
  \includegraphics[width=75mm]{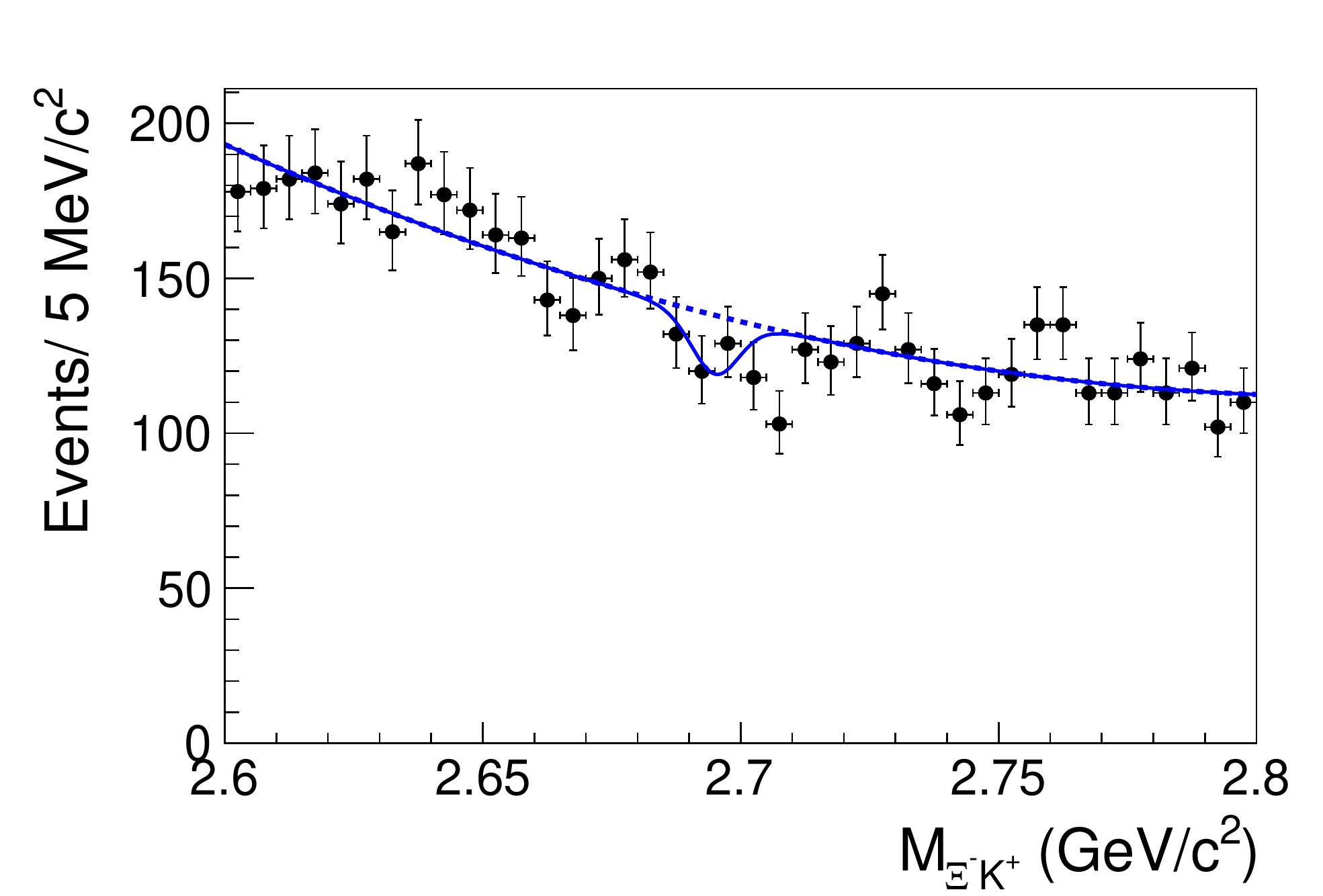}
  \put(-35, 115){\bf (b)}
  
  \includegraphics[width=75mm]{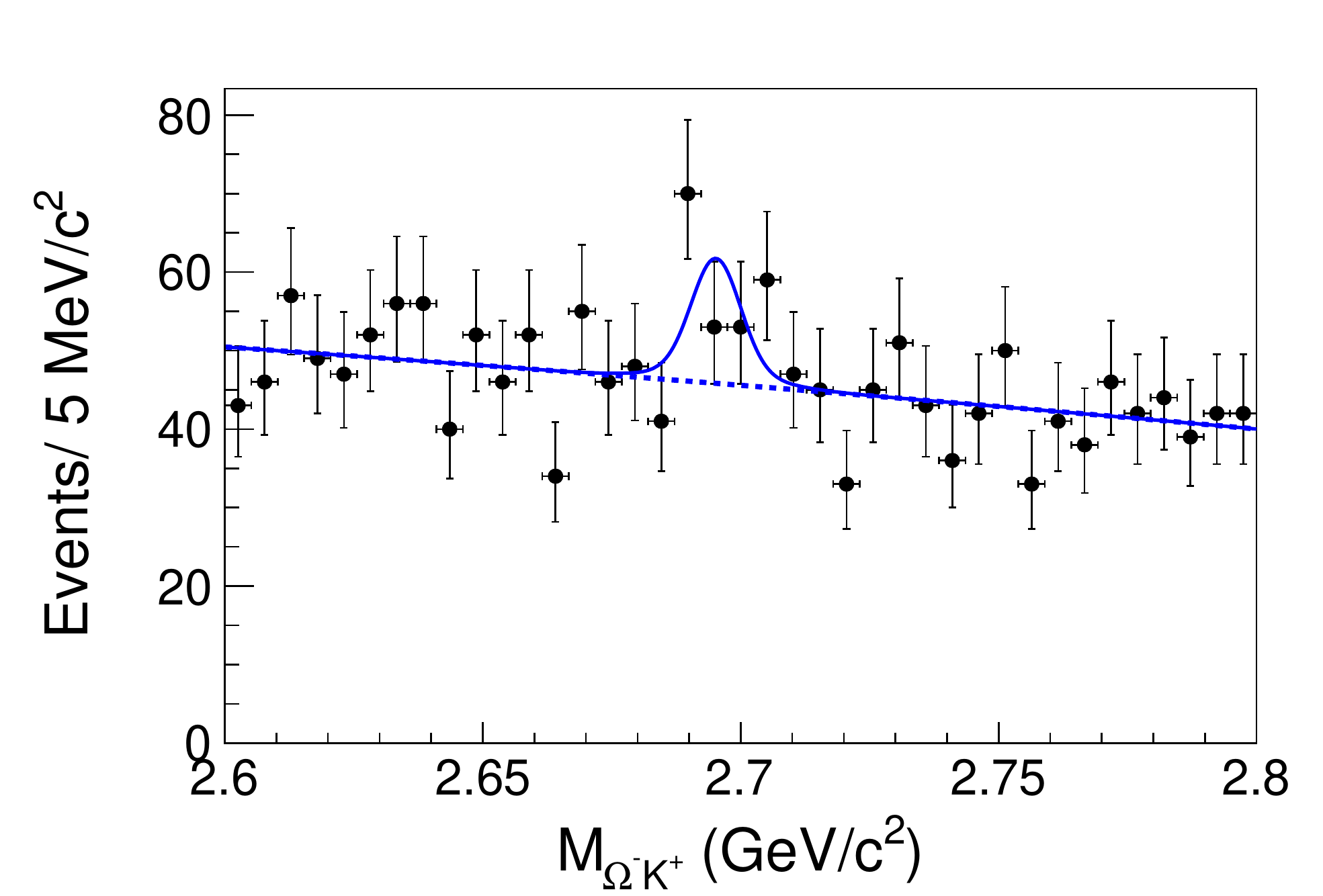}
  \put(-35, 115){\bf (c)}
  \includegraphics[width=75mm]{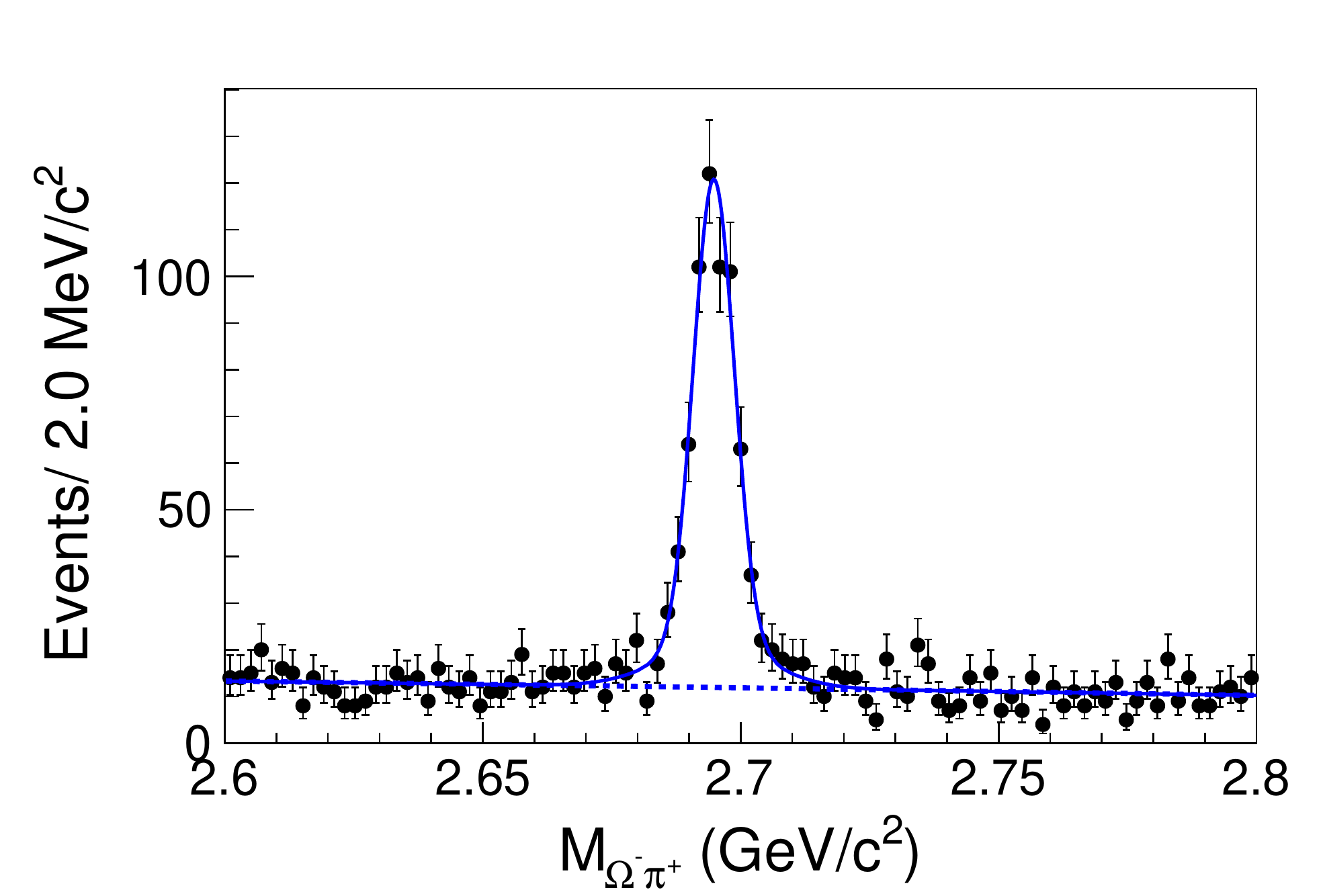}
  \put(-35, 115){\bf (d)}
  \caption{The invariant mass distributions of (a) $\Xi^-\pi^+ $, (b) $\Xi^-K^+ $, (c) $\Omega^-K^+ $, and (d) $\Omega^-\pi^+$ from data samples. The solid curves are the best fits, and the dashed curves are the fitted backgrounds.}
  \label{fig:datafit}
\end{figure}

According to generic simulated samples, no peaking backgrounds are found in the $ \omec $ region for all of the studied modes.
The invariant mass distributions of $\Xi^-\pi^+ $, $\Xi^-K^+ $, $\Omega^-K^+ $, and $\Omega^-\pi^+$ are shown in figure~\ref{fig:datafit}.
To extract the $\omec$ signal yields from each mode, we perform binned maximum-likelihood fits to the $M(\Xi^-\pi^+)$, $M(\Xi^-K^+)$, $M(\Omega^-K^+)$, and $M(\Omega^-\pi^+)$ distributions.
The signal shapes of the $\omec$ are described by double-Gaussian functions, and second-order polynomial functions represent backgrounds.~Considering the central values of two Gaussians are the same, the mass resolution from the double-Gaussian function is characterized by $\sigma$ = $\sqrt{f_1\sigma^2_1+(1-f_1)\sigma^2_2}$, where $\sigma_1$ and $\sigma_2$ are the widths of the first and second Gaussians, and $f_1$ is the fraction of the first Gaussian function.

The mass resolutions for $\Omega^0_c\to\Xi^-\pi^+$, $\Xi^-K^+$, $\Omega^-K^+$, and $\Omega^-\pi^+$ are fixed from the fits to the corresponding simulated signal distributions, which are listed in table~\ref{tabparas}. 
For $\Omega^0_c\to\Xi^-\pi^+ $ and $\Omega^-\pi^+$, the central values of double Gaussian functions are floated since the large uncertainty of $\Omega^0_c$ mass from the world-average (nominal) result~\cite{PDG}. From table~\ref{tabparas}, the central values of the double Gaussian functions in $\Omega^0_c\to\Omega^-\pi^+$ and $ \octxipi $ from the fits are consistent with the nominal mass of the $\Omega^0_c$ within $\pm1\sigma$ and $ \pm2\sigma $, respectively.
For $\Omega^0_c\to\Xi^-K^+$ and $\Omega^-K^+$ which have no significant signals, the central values of the double Gaussian functions are fixed at the nominal mass of $\Omega^0_c$~\cite{PDG}. The parameters of the polynomial functions for backgrounds are free in the fits. 
The fit quality indicated by $\chi^2/{\rm ndf}$ is listed in table~\ref{tabparas}, where $\rm ndf$ is the number of degrees of freedom.

The signal yields and statistical significances are listed in table~\ref{tabparas}. The statistical significances of the $ \omec $ signals are calculated using  $-2\ln\left(\mathcal{L}_{0}/\mathcal{L}_{\max }\right)$, where $\mathcal{L}_{0}$ and $\mathcal{L}_{\max}$ are the likelihoods of the fits without and with $\omec $ signal, respectively~\cite{60}, taking into account the difference in the number of degrees of freedom. 

\begin{table}[htpb]
\centering
\caption{The results of fits to the data, where $N^{\rm fit}$ represents the signal yield and $\Sigma$ is the statistical significance. The uncertainties shown are statistical only.}
\vspace{0.2cm}
\label{tabparas}
\begin{tabular}{c c c c c c}
\hline\hline
Mode & Mass (MeV/$c^2$) & Resolution (MeV/$c^2$) & ~~~$N^{\rm fit}$~~~ & ~~~$\chi^2/{\rm ndf}$~~~& $\Sigma$($\sigma$) \\\hline
$\Omega^0_c\to \Xi^-\pi^+$     & $2692.0\pm1.2$         & $6.6$ (fixed) & $208\pm41$ & $ 0.75 $& $5.1$  \\
$\Omega^0_c\to \Xi^- K^+$      & $2695.2$ (fixed) & $6.3$ (fixed) & $-47\pm23$       & $ 0.88 $& $-$  \\
$\Omega^0_c\to \Omega^- K^+$   & $2695.2$ (fixed) & $6.4$ (fixed) & $41\pm17$        & $ 0.79 $& $2.2$  \\
$\Omega^0_c\to \Omega^- \pi^+$ & $2694.8\pm0.2$         & $6.7$ (fixed) & $606\pm29$ & $ 0.74 $& $>10.0$ \\
\hline\hline
\end{tabular}                                                                                     
\end{table}

By using the normalization mode $\octopi$, we calculate the ratios of branching fractions according to the equations
\begin{equation} \label{eq:1}
  \frac
  {
    \mathcal{B}
    \left(\Omega_{c}^{0} \to \Xi^{-} \pi ^{+} \right)
  }
  {
    \mathcal{B}\left(\Omega_{c}^{0} \to \Omega^{-} \pi^{+}\right)
  }
  =\frac
  {
    N_{\Xi \pi} \cdot \varepsilon_{\Omega \pi} 
    \cdot
    \mathcal{B}\left(\Omega^{-} \to \Lambda K^{-}\right)
  }
  {
    N_{\Omega \pi} \cdot \varepsilon_{\Xi \pi}
    \cdot
    \mathcal{B}\left(\Xi^{-} \to \Lambda \pi^{-}\right)
  },
\end{equation}
\begin{equation} \label{eq:2}
  \frac
  {
    \mathcal{B}
    \left(\Omega_{c}^{0} \to \Xi^{-} K ^{+} \right)
  }
  {
    \mathcal{B}\left(\Omega_{c}^{0} \to \Omega^{-} \pi^{+}\right)
  }
  =\frac
  {
    N_{\Xi K} \cdot \varepsilon_{\Omega \pi}
    \cdot
    \mathcal{B}\left(\Omega^{-} \to \Lambda K^{-}\right)
  }
  {
    N_{\Omega \pi} \cdot \varepsilon_{\Xi K}
    \cdot
    \mathcal{B}\left(\Xi^{-} \to \Lambda \pi^{-}\right)
  },
\end{equation}
and
\begin{equation} \label{eq:3}
  \frac
  {
    \mathcal{B}
    \left(\Omega_{c}^{0} \to \Omega^{-} K ^{+} \right)
  }
  {
    \mathcal{B}\left(\Omega_{c}^{0} \to \Omega^{-} \pi^{+}\right)
  }
  =\frac
  {
    N_{\Omega K} \cdot \varepsilon_{\Omega \pi}
  }
  {
    N_{\Omega \pi} \cdot \varepsilon_{\Omega K}
  }.
\end{equation}
Here, $N_{\Xi \pi}$, $N_{\Xi K}$, $N_{\Omega K}$, and $N_{\Omega \pi}$ are the fitted signal yields for $\Omega^0_c\to\Xi^-\pi^+$, $\Omega^0_c\to\Xi^-K^+$, $\Omega^0_c\to\Omega^-K^+$, and $\Omega^0_c\to\Omega^-\pi^+$,
and $\varepsilon_{\Xi \pi} = 10.7\%$, $\varepsilon_{\Xi K} = 6.1\%$, $\varepsilon_{\Omega K} = 4.7\%$, and $\varepsilon_{\Omega \pi} = 11.6\%$ 
are reconstruction efficiencies. 
These efficiencies are determined as numbers of MC signal events after applying the selection criteria divided by the numbers of generated events. The statistical uncertainties in the determination of efficiencies for the various decay modes are equal to 1\%. They are considered in the systematic estimations, as reported later in this paper.
The branching fractions $\BR (\Xi^{-} \to \Lambda \pi^{-})$ and $\BR (\Omega^{-} \to \Lambda K^{-})$ are (99.887$\pm$0.035)\% and (67.8$\pm$0.7)\%, respectively~\cite{PDG}.

For $\Omega_{c}^{0} \to \Xi^{-} \pi ^{+} $, using the values above, the ratio of branching fractions to the normalization mode of $\Omega^0_c\to\Omega^-\pi^+$ is measured to be
\begin{equation}\label{eq:4}
\frac{\BR(\Omega_{c}^{0} \to \Xi^{-} \pi ^{+} )}{\BR(\Omega_c^0\to \Omega^-\pi^+)}=\left[25.3\pm 5.2({\rm stat.})\right]\%.
\end{equation}

For $\Omega^0_c\to\Xi^-K^+$ and $\Omega^0_c\to\Omega^-K^+$, since the signal significances are less than $3\sigma$, we compute 90\% confidence level (C.L.) upper limits $x^{\rm UL}$ on the signal yields and branching fraction ratios by solving the equation $\int _0^{x^{\rm UL}}{\cal L}(x)dx/\int _0^{+\infty}{\cal L}(x)dx$ = 0.90, where $x$ is the assumed signal yield or branching fraction ratio, and ${\cal L}(x)$ is the corresponding maximized likelihood of the fit to the assumption.

\section{Systematic Uncertainties}

Systematic uncertainties on the branching fraction ratios are summarized in table~\ref{sys}.
The sources of uncertainty are reconstruction efficiency, branching fractions of intermediate states, the statistical uncertainty in the determination of efficiency, the generator model, $\Omega^0_c$ resonance parameters, the uncertainty associated with the fitting procedure, the statistical uncertainty of signal yield in the reference mode of $\Omega^0_c\to\Omega^-\pi^+$. 

The reconstruction-efficiency-related uncertainties include those from tracking efficiency and the PID efficiency.
The uncertainties from tracking efficiency and part of the PID uncertainties are canceled in the ratio to the normalization mode $\octopi$.
The uncertainties in PID are studied via low-background sample of $D^{*+}\to D^0(\to K^-\pi^+)\pi^+$ for charged kaons and pions. The studies show uncertainties of 1.1\% for each charged kaon and 0.9\% for each charged pion.
The systematic uncertainties from the same particles are added linearly, considering a total correlation, while the ones from different particles are summed in quadrature, as there is no correlation.

As the $\Omega^-$ branching fraction uncertainty is cancelled in the ratio to the normalization mode,
only uncertainties of $\BR(\Xi^- \to \Lambda  \pi^-)$ (0.035\%) and $\BR( \Omega ^-\to \Lambda K^-)$ (1.0\%)~\cite{PDG} are included for the $\octxipi$ and $\octxik$ modes.
Using simulated signal events of all the decay modes, the statistical uncertainty in the reconstruction efficiency can be calculated as $\Delta_{\varepsilon}=\sqrt{\varepsilon(1-\varepsilon)/N}$, where $\varepsilon$ is the reconstruction efficiency after all the event selections, and $N$ is the total number of generated events. 
The fractional uncertainty $\Delta_{\varepsilon}/\varepsilon$ is less than 1.0\% in all modes.
Simulated $\Omega_c^0$ decays are generated by the phase space model.
To estimate the uncertainties from MC modeling of the signals, signal MC samples are also generated with an angular distribution of $1-{\rm cos}^2\theta$ or $1+{\rm cos}^2\theta$ at MC-generation level,
where $ \theta $ is the 
angle between the $\Xi^-$ 
or $ \Omega^- $
momentum vector 
and boost direction
of the $\Omega^0_c$ from the laboratory frame in the $ \omec $ rest frame.
The largest differences on the efficiencies between the phase space and $1\pm{\rm cos}^2\theta$ are 7.6\%, 11.2\%, and 5.0\% for $\octxipi$, $\Xi^-K^+$, and $\Omega^-K^+$, and these are are included in uncertainties due to the generator model.

In fitting to the $M(\Xi^-\pi^+)$ and $M(\Omega^-\pi^+)$ distributions, we enlarge the mass resolution by $ 0.2~{\rm MeV}/c^2 $ as indicated by the differences in the mass resolutions between signal MC and data in $ \octxipi $ and $ \Omega^-\pi^+ $ modes and take the difference of the signal yield as the systematic uncertainty of the mass resolution. And for $M(\Xi^-K^+)$ and $M(\Omega^-K^+)$ distributions, we enlarge the mass resolution by $10\%$ since the MC simulation is known to reproduce the resolution of mass peaks within 10\% over a large number of different systems.
In fitting to the $M(\Xi^-\pi^+) $ distribution, we change the $ \omec $ mass to the nominal value~\cite{PDG} and take the difference of the signal yield as the systematic uncertainty of mass central value.
In fitting to the $M(\Xi^-K^+)$ and $M(\Omega^-K^+)$ distributions, we change the $ \omec $ mass by $\pm1\sigma$. The total systematic uncertainty due to $\Omega^0_c$ resonance parameters is obtained by summing the uncertainties of resolution and mass in quadrature for $ \octxipi $. 
We estimate the systematic uncertainties associated with the fitting procedure by changing the order of the background polynomial, the range of the fit, and the number of bins, and take the deviations of signal yields from the nominal fitted results as systematic uncertainties. The total uncertainty is obtained by summing the uncertainties from $\Omega_{c}^{0} \to \Xi^{-} \pi ^{+}$ and the reference mode of $\Omega_{c}^{0}\to \Omega^-\pi^+$ in quadrature. 

The statistical uncertainty of the fitted signal yield for the reference mode ($N_{\Omega^-\pi^+}=606\pm29)$ is 4.8\%.
The statistical uncertainty of the fitted signal yield for the signal mode ($N_{\Xi^-\pi^+}=208\pm41)$ is 19.7\%.
We add them in quadrature to obtain the total statistical uncertainties.

Finally, assuming all the sources are independent and adding them in quadrature, the total uncertainties on the branching fraction ratio measurements are calculated and these are listed in table~\ref{sys}.

\begin{table}[htpb]
\centering
\caption{Relative uncertainties on branching fraction ratio measurements ($\%$). The $\sigma^{\rm syst}_{\rm eff}$, $\sigma^{\rm syst}_{\rm MC}$, $\sigma^{\rm syst}_{\rm GM} $, $\sigma^{\rm syst}_{\BR}$, $\sigma^{\rm syst}_{\rm resonance}$, and $\sigma^{\rm syst}_{\rm fit}$ denote uncertainties from reconstruction efficiency, the statistical uncertainty in the determination of efficiency, generator model, branching fractions of intermediate states, $\Omega^0_c$ resonance parameters, and fitting procedure, respectively. The $\sigma^{\rm stat}_{\rm signal}$ and $\sigma^{\rm stat}_{\rm ref}$ represent the statistical uncertainties in the signal yields from the signal mode and reference mode $\Omega^0_c\to\Omega^-\pi^+$. The $\sigma^{\rm syst}_{\rm sum}$, $\sigma^{\rm stat}_{\rm sum}$, and $\sigma_{\rm sum}$ are total systematic uncertainty, total statistical uncertainty, and total uncertainty, respectively.}
\vspace{0.2cm}
\label{sys}
\begin{tabular}{l|cccccc|c|cc|c|c}
\hline
Decay Mode & $\sigma^{\rm syst}_{\rm eff}$  & $\sigma^{\rm syst}_{\rm MC}$ & $\sigma^{\rm syst}_{\rm GM}$  & $\sigma^{\rm syst}_{\BR}$   & $\sigma^{\rm syst}_{\rm resonance}$   & $\sigma^{\rm syst}_{\rm fit}$  & $\sigma^{\rm syst}_{\rm sum}$   & $\sigma^{\rm stat}_{\rm signal}$ & $\sigma^{\rm stat}_{\rm ref}$ & $\sigma^{\rm stat}_{\rm sum}$ & $\sigma_{\rm sum}$\\\hline
$\octxipi$    &       2.9        &     1.0    &  7.6   &     1.1    &   6.8      &    4.8     &   11.7 & 19.7 & 4.8 & 20.3 & 23.4\\
$\octxik$     &       2.8        &     1.0    & 11.2   &     1.1    &        -   &        -   &   11.6     & - & 4.8 & 4.8 & 12.6\\
$\octok$      &       1.5        &     1.0    &  5.0   &     -      &        -   &        -   &   5.3      & - & 4.8 & 4.8 & 7.2\\
\hline
 \end{tabular}
\end{table}

To estimate the signal significance of the
$\Omega^0_c\to\Xi^-\pi^+$ decay after considering the systematic uncertainties,
alternative fits to the $\Xi^-\pi^+$ mass spectrum are performed by: 
(a) using a first-order or third-order polynomial for background shape; 
(b) enlarging the $ \omec $  mass resolution by  0.2 ${\rm MeV}/c^2$; and 
(c) changing the $ \omec $ mass to the nominal value~\cite{PDG}.
The $ \Omega_c^0 $  signal significance is larger than $4.5 \sigma$ in all cases.
The ratio of branching fractions to the normalization mode of $\Omega^0_c\to\Omega^-\pi^+$ is measured to be
\begin{equation}\label{eq:4}
\frac{\BR(\Omega_{c}^{0} \to \Xi^{-} \pi ^{+} )}{\BR(\Omega_c^0\to \Omega^-\pi^+)}=\left[25.3\pm 5.2({\rm stat.})\pm 3.0({\rm syst.})\right]\%.
\end{equation}

In the calculations of upper limits, the systematic uncertainties are taken into account in two steps. 
First, when we study the additive systematic uncertainties from resonance parameters and the fitting procedure described above, we calculate
the upper limit for each possible case, and take the most conservative upper limit at $90\%$ C.L. on the
number of signal events. 
Then, the likelihood with the most conservative upper limit
is convolved with a Gaussian function whose width is
equal to the corresponding total uncertainty ($\sigma_{\rm sum}$) summarized in Table~\ref{sys}.
The 90\% C.L. upper limits of the signal yields for $ \octxik $ and $ \Omega^-K^+ $ modes are $ 33 $ and $ 70 $ including systematic uncertainties.
And the upper limits at 90\% C.L. on the ratios of branching fractions are 
\begin{equation}\label{eq:5}
\frac{\BR(\Omega_{c}^{0} \to \Xi^{-} K ^{+} )}{\BR(\Omega_c^0\to \Omega^-\pi^+)} < 0.070
\end{equation}
and
\begin{equation}\label{eq:6}
\frac{\BR(\Omega_{c}^{0} \to \Omega^{-} K ^{+} )}{\BR(\Omega_c^0\to \Omega^-\pi^+)}< 0.29.
\end{equation}

\section{Summary}

We have searched for the SCS decays $\octxipi$ and $\Omega^-K^+$ and the DCS decay $\octxik$ for the first time. We report the first evidence of $\octxipi$ with a signal significance of 4.5$\sigma$ including systematic uncertainties. The ratio of branching fractions to the normalization mode of $\octopi$ is $0.253\pm 0.052({\rm stat.})\pm0.030({\rm syst.})$, which is 2.4$\sigma$ away from the theoretical calculation of 0.104 using the CA and pole methods~\cite{094033}, and 
larger than the theoretical calculation of $1.96\times10^{-3}$ using the LFQM method~\cite{093101}.
No significant signals are found in $\octxik$ and $\octok$.
The upper limit at 90\% C.L. on the ratio of branching fractions for $\octxik$ is $0.070$, which is consistent with the theoretical predictions of $1.1\times10^{-2}$ using the CA and pole methods~\cite{094033} and $1.7\times10^{-4}$ using the LFQM method~\cite{093101}.
The upper limit at 90\% C.L. on the ratio of branching fractions for $\octok$ is $ 0.29 $.

\acknowledgments

This work, based on data collected using the Belle detector, which was
operated until June 2010, was supported by 
the Ministry of Education, Culture, Sports, Science, and
Technology (MEXT) of Japan, the Japan Society for the 
Promotion of Science (JSPS), and the Tau-Lepton Physics 
Research Center of Nagoya University; 
the Australian Research Council including grants
DP180102629, 
DP170102389, 
DP170102204, 
DE220100462, 
DP150103061, 
FT130100303; 
Austrian Federal Ministry of Education, Science and Research (FWF) and
FWF Austrian Science Fund No.~P~31361-N36;
the National Natural Science Foundation of China under Contracts
No.~11675166,  
No.~11705209;  
No.~11975076;  
No.~12135005;  
No.~12175041;  
No.~12161141008; 
Key Research Program of Frontier Sciences, Chinese Academy of Sciences (CAS), Grant No.~QYZDJ-SSW-SLH011; 
Project ZR2022JQ02 supported by Shandong Provincial Natural Science Foundation;
the Ministry of Education, Youth and Sports of the Czech
Republic under Contract No.~LTT17020;
the Czech Science Foundation Grant No. 22-18469S;
Horizon 2020 ERC Advanced Grant No.~884719 and ERC Starting Grant No.~947006 ``InterLeptons'' (European Union);
the Carl Zeiss Foundation, the Deutsche Forschungsgemeinschaft, the
Excellence Cluster Universe, and the VolkswagenStiftung;
the Department of Atomic Energy (Project Identification No. RTI 4002) and the Department of Science and Technology of India; 
the Istituto Nazionale di Fisica Nucleare of Italy; 
National Research Foundation (NRF) of Korea Grant
Nos.~2016R1\-D1A1B\-02012900, 2018R1\-A2B\-3003643,
2018R1\-A6A1A\-06024970, RS\-2022\-00197659,
2019R1\-I1A3A\-01058933, 2021R1\-A6A1A\-03043957,
2021R1\-F1A\-1060423, 2021R1\-F1A\-1064008, 2022R1\-A2C\-1003993;
Radiation Science Research Institute, Foreign Large-size Research Facility Application Supporting project, the Global Science Experimental Data Hub Center of the Korea Institute of Science and Technology Information and KREONET/GLORIAD;
the Polish Ministry of Science and Higher Education and 
the National Science Center;
the Ministry of Science and Higher Education of the Russian Federation, Agreement 14.W03.31.0026, 
and the HSE University Basic Research Program, Moscow; 
University of Tabuk research grants
S-1440-0321, S-0256-1438, and S-0280-1439 (Saudi Arabia);
the Slovenian Research Agency Grant Nos. J1-9124 and P1-0135;
Ikerbasque, Basque Foundation for Science, Spain;
the Swiss National Science Foundation; 
the Ministry of Education and the Ministry of Science and Technology of Taiwan;
and the United States Department of Energy and the National Science Foundation.
These acknowledgements are not to be interpreted as an endorsement of any
statement made by any of our institutes, funding agencies, governments, or
their representatives.
We thank the KEKB group for the excellent operation of the
accelerator; the KEK cryogenics group for the efficient
operation of the solenoid; and the KEK computer group and the Pacific Northwest National
Laboratory (PNNL) Environmental Molecular Sciences Laboratory (EMSL)
computing group for strong computing support; and the National
Institute of Informatics, and Science Information NETwork 6 (SINET6) for
valuable network support.

\renewcommand{\baselinestretch}{1.2}

\end{document}